\documentclass[%
 reprint, 
superscriptaddress,
nofootinbib,
 amsmath,amssymb,
 aps
]{revtex4-2}

\usepackage{wasysym}
\usepackage{hyperref}
\usepackage{lipsum}
\usepackage{csquotes}
\usepackage{graphicx}
\usepackage{dcolumn}
\usepackage{bm}

\usepackage{printlen} 
\usepackage{setspace}
\usepackage[capitalise]{cleveref}
\makeatletter
\AddToHook{cmd/appendix/before}{\def\cref@section@alias{appendix}\def\cref@subsection@alias{appendix}}
\makeatother

\usepackage{physics}
\usepackage{xspace}
\usepackage{xcolor}

\newcommand{\suppmat}{Supplemental Material\xspace}

\newcommand{\suppnotemodel}{Supp.~Note~1\xspace}
\newcommand{\suppnoteanalysis}{Supp.~Note~2\xspace}
\newcommand{\suppnoteadditionalresults}{Supp.~Note~3\xspace}

\newcommand{\beq}{\ensuremath{b^\text{eq}}}

\usepackage{color}

\let\oldchi\chi
\let\chi\undefined
\DeclareRobustCommand{\chi}{{\mathpalette\irchi\relax}}
\newcommand{\irchi}[2]{\raisebox{\depth}{$#1\oldchi$}} 

\newcounter{myfigpanel}[figure]
\newcounter{myfigpanelonly}[figure]

\newcommand{\panelletter}[1]{\refstepcounter{myfigpanel}\label{#1}\refstepcounter{myfigpanelonly}\label{onlyletter:#1}\alph{myfigpanel}}
\newcommand{\panel}[1]{(\protect\panelletter{#1})}

\crefalias{myfigpanel}{figure}
\crefname{myfigpanelonly}{panel}{panels}

\let\origcaption\caption
\let\caption\undefined
\DeclareRobustCommand{\caption}[1]{\origcaption{\protect\setcounter{myfigpanel}{0}\protect\setcounter{myfigpanelonly}{0}#1}}

\newcommand{\mycaption}[1]{\caption{\setcounter{myfigpanel}{0}#1}}

\usepackage{ifthen}

\newboolean{showpartopic}

\newcommand{\partopic}[1]{%
  \ifthenelse{\boolean{showpartopic}}{\textbf{#1}}{}%
}

\newcommand{\MPIDS}{\affiliation{Max Planck Institute for Dynamics and Self-Organization, Göttingen, Germany}}
\newcommand{\IDCS}{\affiliation{Institute for the Dynamics of Complex Systems, Göttingen University, Göttingen, Germany}}

\begin{document}

\title{Dimensionality and confinement reshape competition in\\cellular renewing active matter}

\author{Patrick Zimmer}
\MPIDS
\IDCS

\author{Philip Bittihn}
\email{philip.bittihn@ds.mpg.de}
\MPIDS
\IDCS

\author{Yoav G. Pollack}
\email{yoavpol@gmail.com}
\MPIDS
\IDCS

\begin{abstract}
Cellular renewing active matter---assemblies of proliferating and apoptotic cells---underlies tissue homeostasis, morphogenesis, and clonal competition. Previous work in one-dimensional periodic systems identified a fitness advantage associated with rapid dead-cell clearance, an “opportunistic” competition mechanism. Extending this framework, we study two-dimensional cellular aggregates and show that dimensionality modifies the interplay between competition mechanisms for clones with different clearance rates: in 2D, opportunistic and homeostatic-pressure-based competition jointly shape clonal selection, to varying degrees. We then introduce an explicit circular confinement to probe how boundaries modulate this interplay. While opportunistic competition persists, distinct timescale-dependent behaviors emerge through weakened homeostatic-pressure-based competition near boundaries. Structural analysis reveals that confinement promotes tangential alignment and spatially heterogeneous homeostatic pressure, thereby reshaping competitive outcomes at tissue edges. Our study connects newly discovered competition mechanisms with more realistic biological contexts, highlighting how dimensionality and spatial constraints influence tissue structures and modulate competition in heterogeneous cell populations, with implications for tumor growth dynamics and tissue development.\end{abstract}

\maketitle

\section{Introduction}
\label{sec:intro}

The study of active matter—systems composed of units that consume energy at the microscopic scale—has opened new perspectives on the collective behavior of both living and synthetic systems \cite{te2025metareview}. While much work has focused on motile active matter \cite{2020_gompper_2020}, a particularly rich and complex class is proliferating active matter \cite{proliferating_hallatschek_2023}, in which constituents grow and divide. In biological systems such as tissues, this continual proliferation of cells is essential for functionality: enabling fluidization \cite{fluidization_ranft_2010}, self-organization to spatial structures \cite{growing_dellarciprete_2018, isensee2025sensitive} and ordering \cite{IsenseeStressAnisotropy2022}. Combining these with apoptosis and subsequent clearance, continuous turnover and renewal of tissue material underlie both normal homeostasis \cite{banjac2023maintenance,mechanical_shraiman_2005, podewitz2015tissue} and structural changes in morphogenetic events \cite{guillot2013mechanics}. This dynamic turnover plays a complementary role also in competition between different cell populations \cite{basan2009homeostatic, dissipative_basan_2011, pollack2022competitive, de2012drosophila}. Understanding how these characteristics of renewing active matter interact with mechanical forces and spatial organization is central to revealing the physical principles underlying living matter.

In proliferating cellular populations, competition is essential for the suppression of mutated or unfit cells \cite{survival_merino_2016}, especially in the early development of embryos. On the contrary, a competitive advantage of such cells can lead to tumors and cancer. Distinct subpopulations can compete through multiple mechanisms: adversarial competition involves active recognition of opponents via signaling pathways, followed by their elimination by “winner” cells \cite{baker2020emerging,wagstaff2016mechanical,vincent2013mechanisms}. While coexistence is in principle possible, non-adversarial competition enables certain populations to dominate without any killing mechanism or even sensing other cell types, as demonstrated by neutral competition among stem cells, where clonal dominance emerges stochastically without direct cell-cell antagonism \cite{de2012drosophila}. For example, kinetic cellular competition arises from intrinsic differences in proliferation rates, whereby faster-dividing cells progressively outnumber slower ones \cite{baker2020emerging}, i.e., the faster the division, the better. However, in homeostatic tissues, a high division rate is counterbalanced by increased cell death, invalidating the argument for simple kinetic advantages \cite{levayer2020solid}. When growth and death depend on mechanical pressure, and the homeostatic point where they balance differs between subpopulations, homeostatic pressure can determine the outcome \cite{basan2009homeostatic, dissipative_basan_2011}: the subpopulation with the higher homeostatic pressure becomes dominant. While some competition drivers are long-established, new directions explore the impact of adhesion and force transmission ability on tissue competition \cite{Schoenit2025}.

A recent theoretical study by Pollack et al. \cite{pollack2022competitive} established a new competition mechanism in intrinsically homeostatic systems, complementary to that of homeostatic pressure differences. Their work, based on a one-dimensional (1D) agent-based model, demonstrated that clones capable of rapidly eliminating their own dead matter increase their competitive success over those with slower degradation of dead cells. This advantage—termed the “opportunistic competition” mechanism—arises because faster clearance leads to higher population turnover rates, allowing for a quicker response to space being freed. This mechanism was also found to be able to dominate over small differences in homeostatic pressure that favor the opposing clone.

In this work, we extend the findings of Pollack et al. by investigating the dynamics in an epithelial-like, two-dimensional (2D) system and its effects on cellular competition. Our model incorporates different degradation modes, identical to the original study, which vary the removal process of dead cells, ranging from immediate clearance to delayed degradation. First, we characterize the interplay of opportunistic and homeostatic pressure competition in a periodic domain, and compare it to the 1D dynamics. We validate the competition dynamics found by Pollack et al. in 2D for most competition cases, and find notable exceptions, where homeostatic pressure counters opportunistic competition.

Pushing the applicability of the model further, we then turn to consider environmental mechanical constraints, which are invariably induced in real systems, such as for developing embryos or growing tumors \cite{compressive_delarue_2014,shen2020detecting}, where the balance between proliferation and removal is influenced by the surroundings \cite{compressive_delarue_2014, montel2012stress, montel2012isotropic}. Dynamics within tumors, which in themselves can be clonally heterogeneous \cite{gerlinger2012intratumor, spatial_waclaw_2015}, can be affected by their healthy confining tissue, with possible implications for competition taking place within the tumor itself.

Therefore, we also aim to understand the impact of a finite size and an explicit confinement on our cell populations. We model cellular aggregates enclosed within a circular boundary and study the spatial aspects imposed by the confinement to compare them to the periodic bulk-like system. We find that the competitive success of individual clones is influenced by the confining wall, where the dynamics are locally reminiscent of the 1D system.

\section{Model}
\label{sec:model}
\begin{figure*}[ht]
    \centering
    \includegraphics[width=1.0\textwidth]{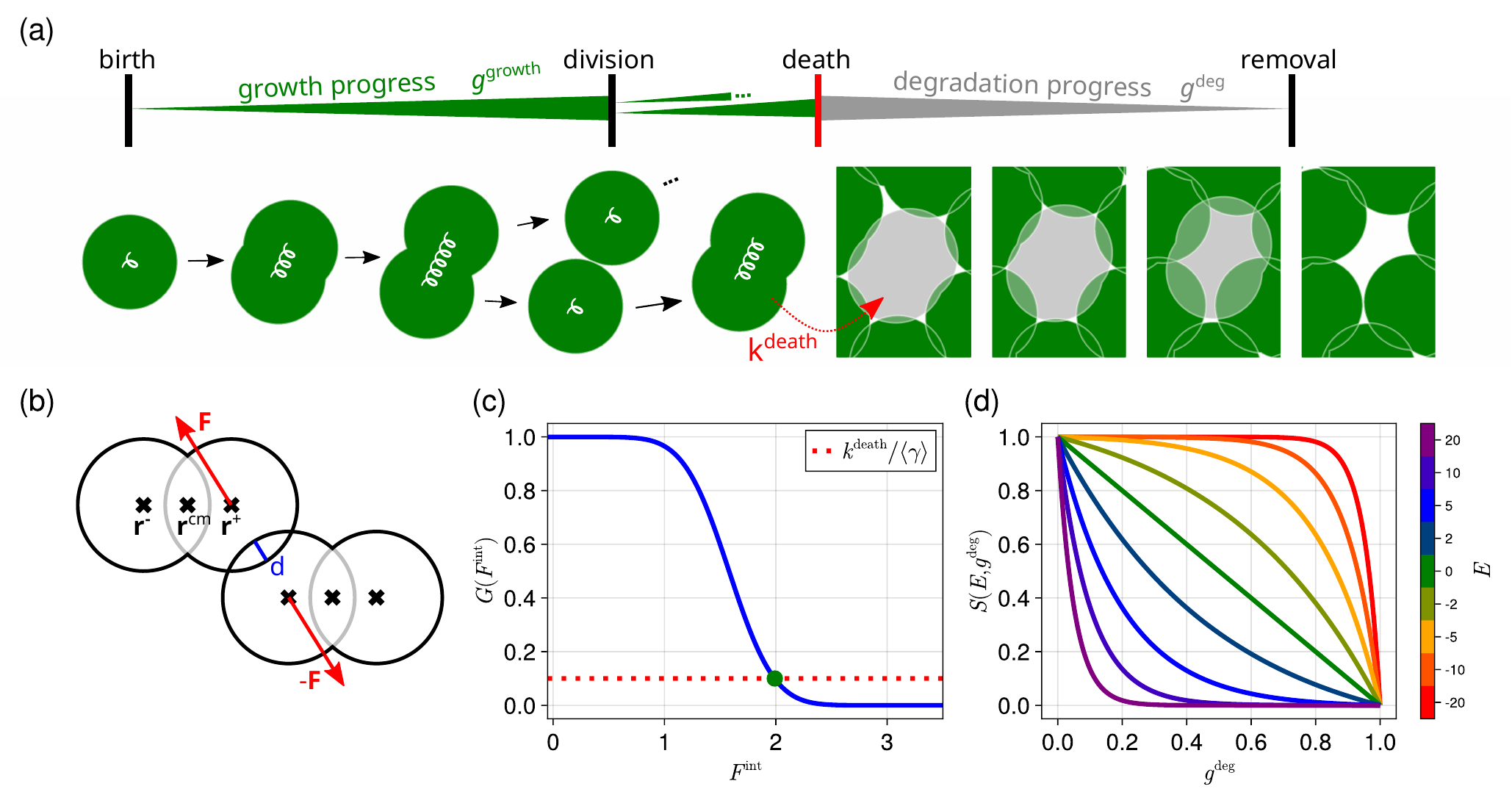}
    \mycaption{2D Agent-based model.
        \panel{pan:cellstages}~Stages in the lifecycle of a dumbbell-shaped cell. Starting from a living cell (green), which can grow and divide for multiple cycles, until the cell dies eventually, stopping all growth. Dead cells degrade until they are removed. Duration of active and passive phase are not to scale.
        \panel{pan:nodeinteraction}~Cell nodes repel each other on contact through Hertzian forces $F$. Forces increase with overlap distance $d$.
        \panel{pan:growthmodulation}~Cell growth can be inhibited when the population is crowded locally, which is implemented by the monotonically decreasing function $G(F^\text{int})$. A cell's probability to die is unaffected by its environment. Cross-section (green) marks a theoretical homeostasis point, at which growth and death balance for a stable homeostasis.
        \panel{pan:scalingfactor}~The scaling factor $S$ linearly affects interaction forces of dead cells with their neighbors during their degradation. The degradation factor $E$ modulates the force scaling. At high negative $E$ (red), dead cells stay structually intact during degradation, while cells with high positive $E$ (purple) lose their repellent force quickly. For small $|E|$ (green), the degradation is gradual. Low scaling factors allow for higher overlap to neighboring cells. Alive cells have $S=1$.}
	\label{fig:figure1}
\end{figure*}

The cell model used in this study captures multiple phenomena seen in real cells—from growth and division up to death and degradation—all in an idealized manner. The progression throughout the cell's life and death is illustrated in \cref{pan:cellstages}, with representative visualizations of the cells' states below each stage. The relative duration of the cells' active and passive phases is not shown to scale, to reserve enough space for the sketches of different states. Each cell is a two-dimensional (2D) dumbbell-shaped agent, consisting of two circular nodes of radius $R$ connected by a non-linear internal spring (backbone). A node diameter defines the unit length. An alive cell grows by increasing the backbone's rest length \beq\,with a base growth rate $\gamma$, which causes the nodes to separate. The state in the growth cycle is tracked by the growth progress $g^\text{growth}=\beq/2R$.  Once it reaches one, two new cells replace the mother cell's nodes. This ensures a continuous description of division processes and involved forces \cite{eDLS_DumbbellandRods}. Daughter cells inherit their mother cell's backbone orientation.

Two different cells interact only on contact by repulsive Hertzian forces \cite{hertz_contact}, which are proportional to the overlap $d = |\mathbf{d}|$ in the following manner (see \cref{pan:nodeinteraction})
\begin{equation}
    \mathbf{F} \propto d^{3/2}\mathbf{\hat{d}}.
\end{equation}
Within a cell, forces on nodes by the backbone follow a similar mechanical behavior, only substituting $d$ by the compression of the backbone. This model aims to simulate 2D dynamics, where simple nearest neighbor interactions as used in Ref.~\citenum{pollack2022competitive} are not applicable due to more complex overlap configurations compared to 1D. Therefore, an additional softness factor interpolates between interacting nodes to prevent force discontinuities, as described in Ref.~\citenum{eDLS_DumbbellandRods} and in the \suppmat. Here, we use this 2D-capable model both for 1D and 2D simulations.

In a finite-size system, proliferating cells will eventually fill all available space. To reach a homeostatic state, a systematic feedback needs to be in place that adjusts cell proliferation or removal rates depending on the current state of the system. In our system, removal—which is introduced in more detail below—happens at a constant rate, while locally high densities or crowdedness can inhibit cell growth. An accessible measure for local crowdedness is the mechanical force on the cell's backbone $F^\text{int}$. To slow growth appropriately, the speed at which a cell progresses through its growth cycle is modulated by a function $G(F^\text{int})$ \cite{pollack2022competitive}, such that $\dot{g}^\text{growth} = \gamma G(F^\text{int})$ with
\begin{equation}
    G(F^\text{int}) = 0.5\left(1-\text{erf}\left(\lambda \left(\frac{F^\text{int}}{F^\text{stall}}-1\right)\right)\right).
\end{equation}
Here, $F^\text{stall} = 3.16$, the typical force in a 1D system where cells overlap by $d \approx 0.1$ and growth is completely inhibited ($G \gtrsim 0$). The growth modulation can be seen in \cref{pan:growthmodulation}.

Living (active) cells have no limitation on the number of divisions in our model, but they die with a constant probability per unit time of $k^\text{death} = 0.01$, turning into dead (passive) cells incapable of growth and division. Those dead cells persist in the system for a time $T^\text{deg} = 10$, during which they degrade mechanically by reducing the strength of their repulsive interactions (see \cref{pan:cellstages}, gray degradation stage). The decrease of steric forces is achieved by a scaling factor $S$ \cite{pollack2022competitive}, which decreases monotonically from $1$ to $0$ during the degradation time $T^\text{deg}$ according to the degradation progress $g^\text{deg}(t)=\left(t-t^\text{death}\right)/T^\text{deg}$ (normalized time after death). The time course of this decrease is parameterized by the so-called degradation factor $E$, such that
\begin{equation}
    S(E,g^\text{deg}(t)) = \begin{cases}
    1, & \text{alive}\\
    1-\frac{\left( 1- e^{-E g^\text{deg}(t)}\right)}{1+e^{-E}}, & \text{degrading.}
    \label{eq:scalingfactor}
\end{cases}
\end{equation}
This leads to fast softening for large positive $E$ (see \cref{pan:scalingfactor}, purple line) and slow for large negative $E$ (red), with gradual degradation for $|E|\approx 0$ (green). In the limit $E \rightarrow \pm\infty$, cells instantaneously lose all structure right after death or at the very end of the degradation phase, respectively. Technically, while still present in a simulation, a cell with a sufficiently low scaling factor $S\lesssim0.1$ is unable to interact with neighboring cells, therefore being effectively removed. Hence, we define the effective degradation rate as in Ref.~\citenum{pollack2022competitive}
\begin{align}
    k^\text{deg}(E) & = \frac{1}{t|_{S=0.1}-t|_{S=1}}\nonumber\\
                    & = \frac{E}{T^\text{deg}\ln\left(1-0.9(1-e^{-E}\right))}.
    \label{eq:kdeg}
\end{align}
In this study, a clone or species is completely defined by its degradation factor $E$. All other parameters are kept constant. Clones in competition with one another therefore only differ in their degradation dynamics.

Simulations use the in-house developed, open-source simulation framework \textit{InPartS} \cite{InPartS}, which evolves the degrees of freedom using overdamped dynamics. The \textit{DiskCell} model \cite{eDLS_DumbbellandRods} implements core functionality of steric interactions and proliferation with important implementation details for a 2D model. This base model is expanded by the pressure-sensitive growth and the degradation aspects of the model described in Ref.~\citenum{pollack2022competitive}. Additional details on the implementation and initialization can be found in \suppnotemodel.

\section{Results}
\label{sec:results}

\subsection{Dimensionality effect on competition}
\begin{figure}[ht]
    \centering
    \includegraphics[width=0.5\textwidth]{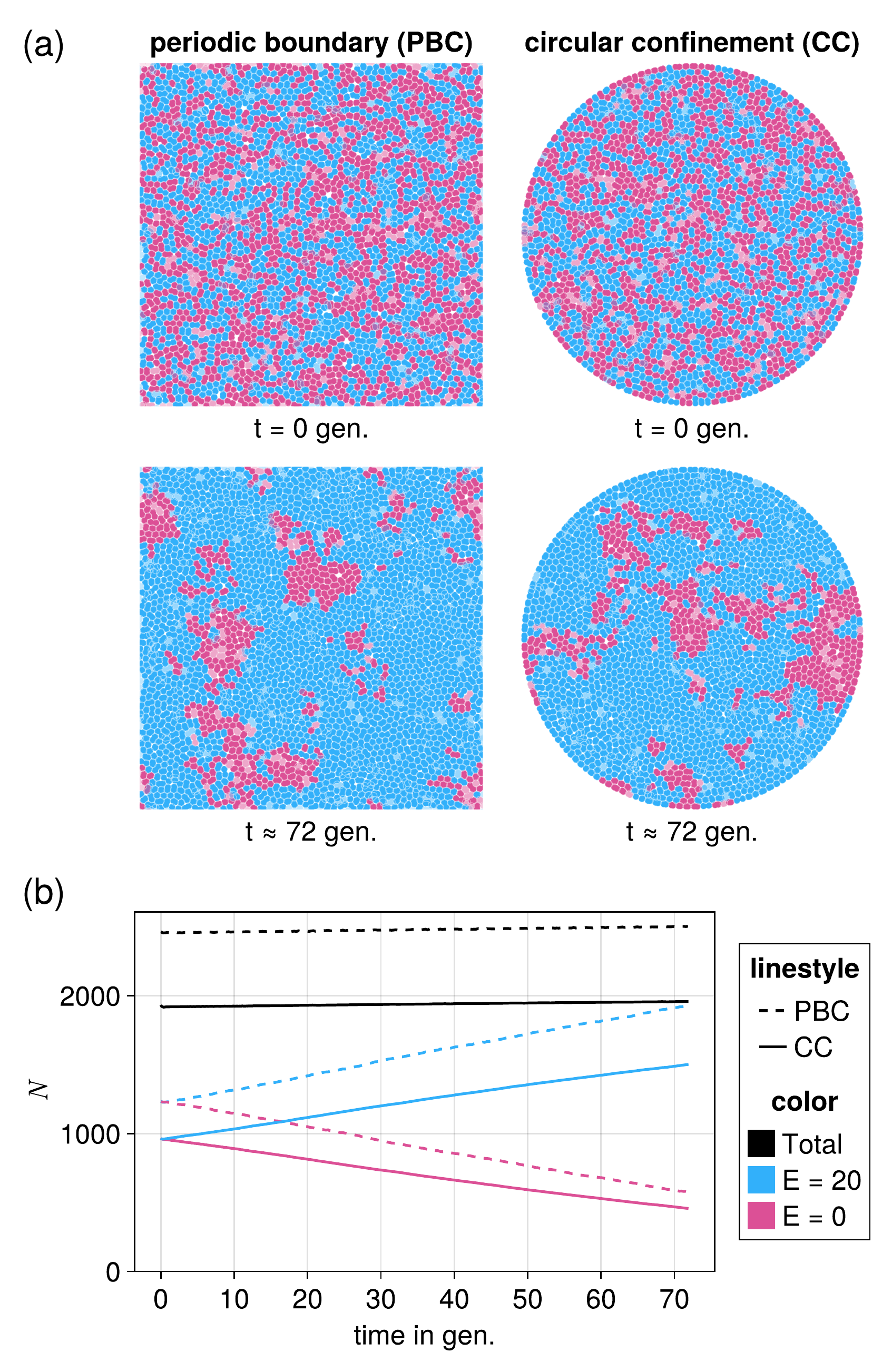}
	\mycaption{Cellular competition in two dimensions (2D)
        \panel{pan:comp_snapshots}~Simulation snapshots in systems with two clones, one with fast dead matter elimination (cyan, $E=20$) and the other with slower and gradual degradation (magenta, $E=0$). Initially well mixed ($t=0\,$gen.) populations in periodic boundary conditions (“PBC”, left) and circular confinement (“CC”, right). The dynamics favor a specific clone over time, having a visibly larger population at the end of the simulation ($t\approx72\,$gen.).
        \panel{pan:comp_global_timeseries}~Time series of global population numbers $N$ in total (black) and of either clone (magenta/cyan). The type of environment is indicated by solid or dashed line for confined population or in periodic system, respectively.}
	\label{fig:figure2}
\end{figure}

Two possible competition mechanisms are examined in this work, which were previously established in Pollack et al. \cite{pollack2022competitive} to be relevant in the scenario of homeostasis under pressure regulation of growth and gradual removal of dead cells: homeostatic pressure and opportunistic competition. While further mechanisms could be at play, we focus here on the interplay between these two.
In the homeostatic pressure mechanism \cite{basan2009homeostatic}, the fitness of a clone is determined by the pressure at which it balances division and death. There, when two clones are placed in competition, the one that stabilizes at a higher pressure expands and presses the other one to a negative division-death imbalance. 

Opportunistic competition, as first described by Pollack et al., arises between clones by differences in the removal dynamics of their own dead cells, which affect the overall fraction of actively growing cells (active density). To see how dead cell removal results in a difference in active cell proportion, we recall that the homeostatic active density $\rho^\text{h}$, i.e., the density of proliferating cells, depends on the degradation mode $E$ in the following way
\begin{equation}
    \rho^\text{h}(E) = \frac{\tilde{\rho}^\text{t}}{1+\frac{k^\text{death}}{k^\text{deg}(E)}},
    \label{eq:activedensity}
\end{equation}
(see Ref.~\citenum{pollack2022competitive}) with total density $\tilde{\rho}^\text{t}$ and the effective degradation rate $k^\text{deg}(E)$ (Eq.~\eqref{eq:kdeg}). Thus, clones that eliminate dead matter more rapidly maintain a larger active fraction, which we confirm later numerically (as seen in \cref{pan:hom_global_activedensity} by the strictly monotonic behavior along the $E$ coordinate following the theory (green line)). Since free space is reclaimed mostly by active cells, a higher active cell proportion (for fast-degrading clones) translates into a fitness advantage acting at interfaces (where both clones compete for the same free space).

First, we study the two competition mechanisms in a domain with periodic boundary conditions (PBC) in one and two dimensions (1D and 2D). The 1D simulations aim to reproduce former results by Pollack et al. \cite{pollack2022competitive} in our 2D-capable model and serve as a foundation for a meaningful comparison between 1D and 2D. The simulation domain in 1D is of length $1000$ (unit length is one node diameter), and in 2D of size $50$x$50$. To later explore how boundaries influence competition, we additionally introduce a simulation geometry with circular confinement (CC) of radius $25$ in 2D. In all scenarios, mixtures of two cell populations are initialized at time $t=0$ in equal proportions, only differing in their degradation mode (degradation factor $E$), and then evolved for $5000$ time units ($\approx72$ generations\footnote{A generation is the average doubling time, which is set by balanced growth matching the death rate, $1\,\text{gen.}=\ln(2)/k^\text{death}$.})---a time in which neither clone would take over the whole population. Representative snapshots illustrate the core expectation of the opportunistic mechanism: the population with faster dead matter removal (cyan, $E=20$) expands at the expense of the slower one (magenta, $E=0$), both in PBC and CC (\cref{pan:comp_snapshots}). Time series of global cell count (alive and dead cells, black) and both clones individually, averaged over $100$ realizations in PBC ($1000$ in CC), are shown in \cref{pan:comp_global_timeseries} (PBC as dashed lines and CC as solid lines). Clearly, the population numbers grow or shrink monotonically, validating the visual trends.

In a competition scenario, the clone with a higher active density $\rho^\text{h}$ is labeled H (high), while the other is labeled L (low). To measure the fitness of a clone H or L, we count its number of cells $N_\text{[H,L]}$ (alive and dead cells) and use the population fraction $N_\text{[H,L]}/N^\text{t}(t_f)$, where $N^\text{t}$ is the total number of cells in the system. We quantify fitness typically from the perspective of clone H, and at the end of the simulation, $t_f \approx 72\,$generations.

\begin{figure}[ht]
    \centering
    \includegraphics[width=0.5\textwidth]{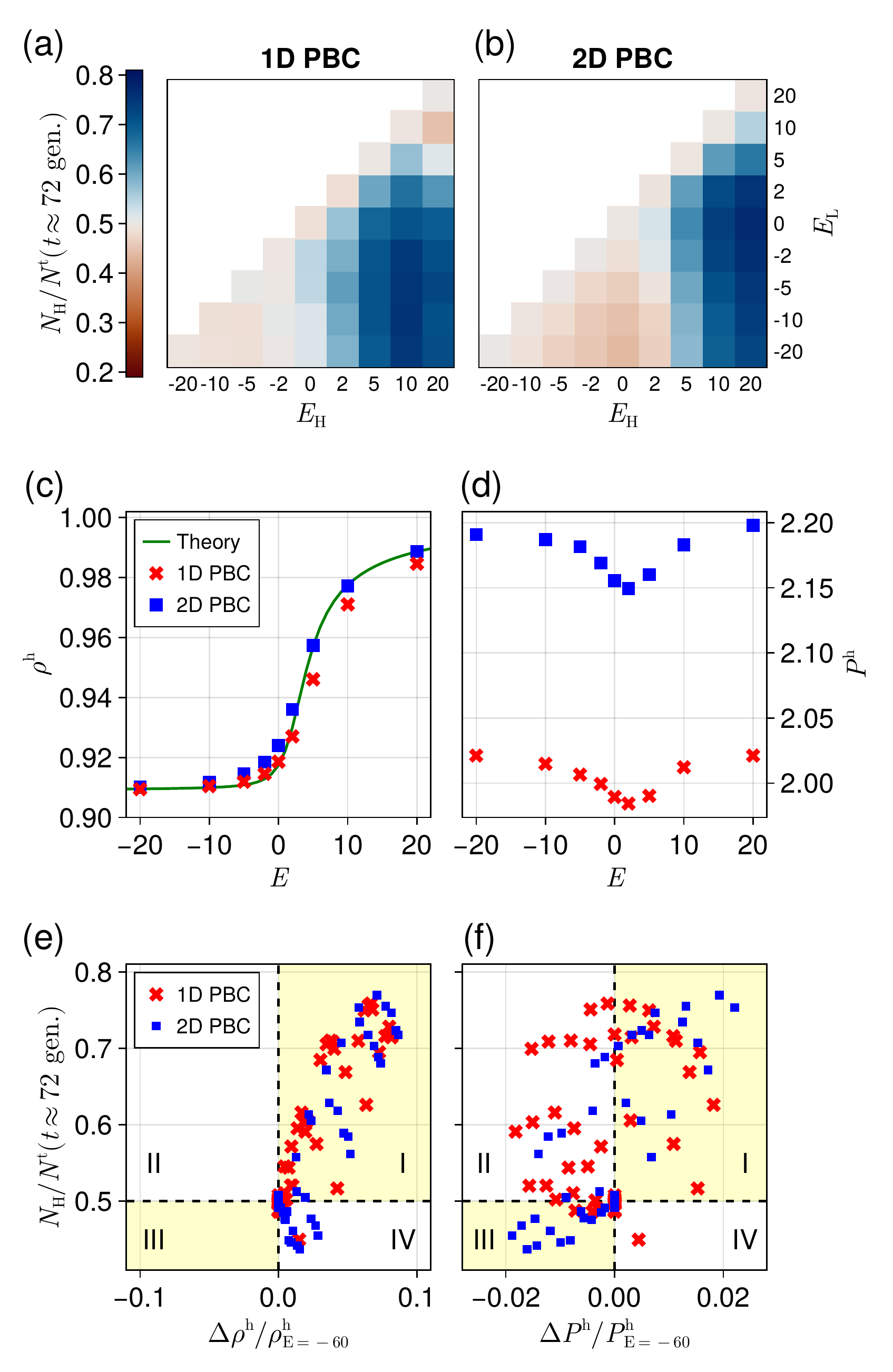}
	\mycaption{Quantification of competition outcomes.
        \panel{pan:comp_global_matrix_1D}~Competition matrix summarizes results for global fitness $N_\text{H}/N^\text{t}$ at the end of the simulation for 1D periodic (PBC) systems. Each data point corresponds to a specific mixture of H and L clones. Global fitness is color-coded, where blue/red means that the H clone increases/decreases in population size.
        \panel{pan:comp_global_matrix_2D}~The same visualization for the 2D PBC system.
        \panel{pan:hom_global_activedensity}~Prediction of homeostatic active density $\rho^h$ according to \eqref{eq:activedensity} (green line) and measurements in homeostasis for a set of degradation factors $E$.
        \panel{pan:hom_global_force}~Homeostatic pressure $P^\text{h}$ in the same two systems.
        \panel{pan:comp_global_correlation_actdens}~Correlation between global fitness and difference in homeostatic active density $\Delta\rho^\text{h}$ (normalized) of competing clones. 
        \panel{pan:comp_global_correlation_pressure}~Correlation of global fitness to normalized difference in homeostatic pressure $\Delta P^\text{h}$.}
	\label{fig:figure3}
\end{figure}

Fitness is first measured globally in the PBC systems in 1D and 2D, for all combinations of competitors with $E \in [-20,-10,-5,-2,0,2,5,10,20]$ (values that sample from slow to fast degradation, within the limits explored by Pollack et al.---full range including extremes $E\mapsto\pm \infty$ see \suppmat), as seen in \cref{pan:comp_global_matrix_1D} and \cref{pan:comp_global_matrix_2D}. Each data point in the competition matrix corresponds to a specific mixture of one H clone (x-axis) and one L clone (y-axis). Global fitness is color-coded, where blue (red) means that the H clone increases (decreases) in population fraction from the initial $50\%$. 
Unsurprisingly, most mixtures follow the prediction of the opportunistic mechanism: clones with higher active density tend to win the competition. However, in 2D, a region of combinations appears where the H clone unexpectedly loses, revealing that active density alone cannot account for all outcomes. Neutral scenarios, where the two populations have identical degradation, show no advantage to either side, as expected.

To separate the contribution of the two mechanisms to fitness, we next characterize each clone in isolation, measuring active density and homeostatic pressure in homogeneous populations of only a single clone. As shown in \cref{pan:hom_global_activedensity}, the homeostatic active density $\rho^\text{h}$ follows the prediction from Eq.~\eqref{eq:activedensity} well and, importantly, recovers the strict monotonic increase with growing degradation factors $E$. In the 1D system, the global average homeostatic pressure $P^\text{h}$ for clones with small $|E|$ is determined by the balance point between growth and death (see \cref{pan:growthmodulation}). For higher $|E|$, fluctuations in pressure increase and so does the mean pressure, as seen in \cref{pan:hom_global_force}\footnote{The link between pressure fluctuations and homeostatic pressure was previously established in Appendix E of Ref.~\citenum{pollack2022competitive}.}. The same relative behavior is also present in 2D, only shifted towards slightly higher pressures.\footnote{Here, the average cell's backbone force serves as a proxy for homeostatic pressure. However, we confirmed that it shows identical behavior to the virial pressure calculated from stress tensor contributions (see \suppnoteadditionalresults).}

To understand which traits act as the primary drivers of the competition, we examine how differences between the two clones' properties are reflected in the final competition outcome. Specifically, we ask: when one clone has a higher active density in homeostasis or a higher homeostatic pressure, does this reliably translate into a competitive advantage?

We answer this by comparing the measured fitness to the difference in active density $\Delta \rho^\text{h} = \rho^\text{h}_\text{H} - \rho^\text{h}_\text{L}>0$, which reflects the opportunistic mechanism, and to the corresponding differences in homeostatic pressure, $\Delta P^\text{h} = P^\text{h}_\text{H} - P^\text{h}_\text{L}$, which reflects the pressure-driven mechanism. We start by providing an intuition for how variations in active density and homeostatic pressure relate to the competition outcome, with a positive (negative) homeostatic quantity $(\Delta\rho^\text{h}, \Delta P^\text{h})$ offering a first indication that the mechanism works in favor of clone H (L). To understand this visually, one can divide the data into four quadrants, where fitness above or below the initial $0.5$ indicates which clone dominated, and each predictor's sign determines the corresponding mechanism's expected direction of advantage.

In 1D, the picture is unambiguous: active density correlates positively with fitness. Red data points in \cref{pan:comp_global_correlation_actdens} cluster predominantly in quadrant I, with only minor noise around the origin. In contrast, homeostatic pressure doesn't show any strong systematic correlation with fitness. Some red points in \cref{pan:comp_global_correlation_pressure} lie in quadrant II, meaning that clone H wins the competition despite having a lower homeostatic pressure. Thus, in 1D, the competition landscape in the considered parameter region is simple: the opportunistic mechanism is largely sufficient to explain the outcome of the competition.

Having confirmed the results of Pollack et al., we now turn to the 2D PBC scenario. In contrast to the simpler behavior in 1D, increasing dimensionality enriches the competition landscape: the opportunistic mechanism is not always dominant anymore (which we already saw in \cref{pan:comp_global_matrix_2D}). Combinations where the H population diminishes (filling quadrant IV in \cref{pan:comp_global_correlation_actdens}) correspond to ones where the homeostatic pressure mechanism favors the L clone (filling quadrant III in \cref{pan:comp_global_correlation_pressure}). Conversely, there are still situations where the homeostatic pressure mechanism alone predicts an L advantage, yet clone H still wins. These cases are similar to the 1D dynamics, where the predictors have opposite signs as well, but the opportunistic driver is strong enough to determine the outcome (filling quadrant II in \cref{pan:comp_global_correlation_pressure}).

Overall, 2D competition therefore reveals a richer interplay between the mechanisms: neither active density nor pressure differences alone fully predict the dynamics, and the relative strength of the two mechanisms becomes decisive whenever their predictions diverge. This stands in marked contrast to the 1D system, where homeostatic pressure never overrides the opportunistic trend in the examined parameter range. Below in Sec.~\ref{subsec:PID}, we carry out a more quantitative comparison of the relative contributions of these two competition drivers.

\subsection{Influence of mechanical confinement}

Having seen that domain dimensionality can influence the competition outcome, the question arises whether other spatial or geometric factors of biological relevance, such as mechanical boundaries, could be decisive as well. Here, we break the homogeneity of space and introduce an explicit circular confinement (CC) as shown in \cref{pan:comp_snapshots}. We then repeat the competition and homeostasis simulations and subsequent analysis on a global scale, and recover results closely following the ones from 2D PBC (see \suppnoteadditionalresults). However, since local structural changes—such as cells aligning to the wall with their long axis—are expected in confined populations, we now investigate whether competition varies spatially. To this end, we observe the population dynamics of subpopulations in radial bins (annuli) from the center towards the confining wall (see \suppnoteanalysis for technical details).

\begin{figure}[ht]
    \centering
    \includegraphics[width=0.50\textwidth]{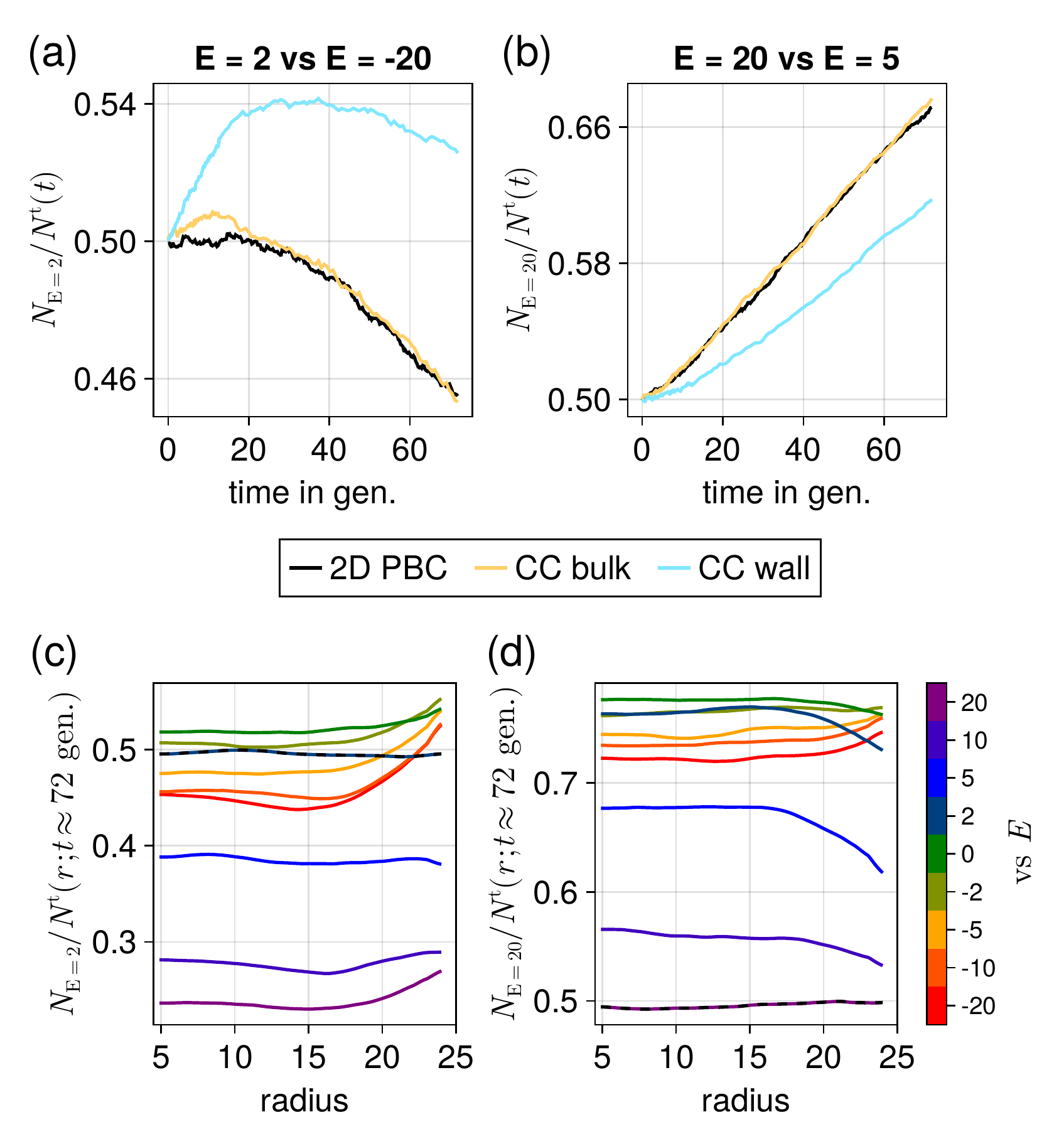}
	\caption{Spatially resolved competition dynamics and fitness; illustrative pairings
        \panel{pan:comp_radial_2vs-20}~Time series of population fraction of clone $E=2$ in competition against a clone with $E=-20$, measured in the center of the population (bulk, orange) and near the confining border (wall, cyan). Global behavior in 2D PBC as a reference (black).
        \panel{pan:comp_radial_20vs5}~Same measurements for clone $E=20$ in competition against $E=5$.
        \panel{pan:comp_radial_fitness_2}~Spatial measurement of fitness of clone $E=2$ in competition scenarios against all other clones (see color bar for adversary clone color). Neutral competition case indicated by a dashed line (examples discussed in text: red: $E=2$ vs $E=-20$, ink blue + dashed black: $E=2$ vs $E=2$).
        \panel{pan:comp_radial_fitness_20}~Same measurements for the clone $E=20$ (with the ink-blue curve corresponding to the same competition scenario as the purple curve in panel c, only viewed from the winner's side).}
	\label{fig:figure4}
\end{figure}

Indeed, we observe altered population dynamics between populations in the bulk and near the border, including some cases where the competition trend is reversed by the presence of the boundary. To illustrate this clearly, we start with such a case, specifically the competition of a clone with $E_\text{H}=2$ (gradual dead matter removal) against a $E_\text{L}=-20$ clone (slow removal). In 2D PBC, the population with $E=2$ is a loser, despite removing dead matter faster than its opponent, as explained above. In \cref{pan:comp_radial_2vs-20}, time series of population fractions $N_{E=2}/N^\text{t}$ are shown measured in the domain center (bulk, orange) and near the wall (cyan). Dynamics in 2D PBC (black) are shown for reference. While in bulk, the dynamics follow roughly the dynamics in PBC, the competition shows qualitatively different dynamics near the confining wall, namely the H clone having a significantly higher population fraction, even above $0.5$, i.e., locally being a winner (at least transiently). Here we also find a timescale dependence and non-monotonic trends in the population fraction: an initial growth period in favor of the H clone (up to $\approx 25$ generations at the wall and $\approx 15\,$gen. in the bulk), after which it shrinks again.

However, the effect of the boundary on competition dynamics is not limited to cases where it causes non-monotonicity, as can be seen in a competition between $E_\text{H} = 20$ vs $E_\text{L} = 5$ (\cref{pan:comp_radial_20vs5}), where the H clone wins as expected from the opportunistic mechanism. In this case, border and bulk dynamics are also different, although less drastically, showing a reduced population growth rate of H at the confining wall, while displaying strictly monotonic growth in all cases. This means that competition strength is reduced near the edge.

The pairwise results naturally raise the question of how each clone fits into the wider hierarchy of competitors. To address this, we compare clone $E=2$ with every other clone across the degradation range, as shown in \cref{pan:comp_radial_fitness_2}. There, we measure the fitness (population fraction at $t\approx72\,$gen.) of that specific clone spatially as a function of radial distance to the colony center. The “sanity check” scenario, where two populations with identical $E$ compete, is expected to show no fitness advantage at all, as validated in the figure (alternating ink blue and black line). Since globally, clone $E=2$ is typically a loser (with a minor advantage against $E\in[-2,0]$), fitness is mostly below $0.5$, but it grows higher near the wall in most cases. This is in alignment with the increased population fraction at the wall seen in \cref{pan:comp_radial_2vs-20}, which here corresponds to the red curve. In this specific case, the increase at the wall is even sufficient to make the $E=2$ the local winner (at the end of the simulation period), as seen from fitness values above $0.5$ at large $r$.

Analogously, fitness of clone $E=20$ in competition against all others is shown in \cref{pan:comp_radial_fitness_20} (with the sanity scenario here being the alternating purple and black line). In this case we observe the opposite behavior: a clone, winning against all other clones globally, tends to have reduced fitness at the wall against some clones. The decreased population fraction at the wall seen in \cref{pan:comp_radial_20vs5} here corresponds to the blue curve.

\begin{figure}[ht]
    \centering
    \includegraphics[width=0.5\textwidth]{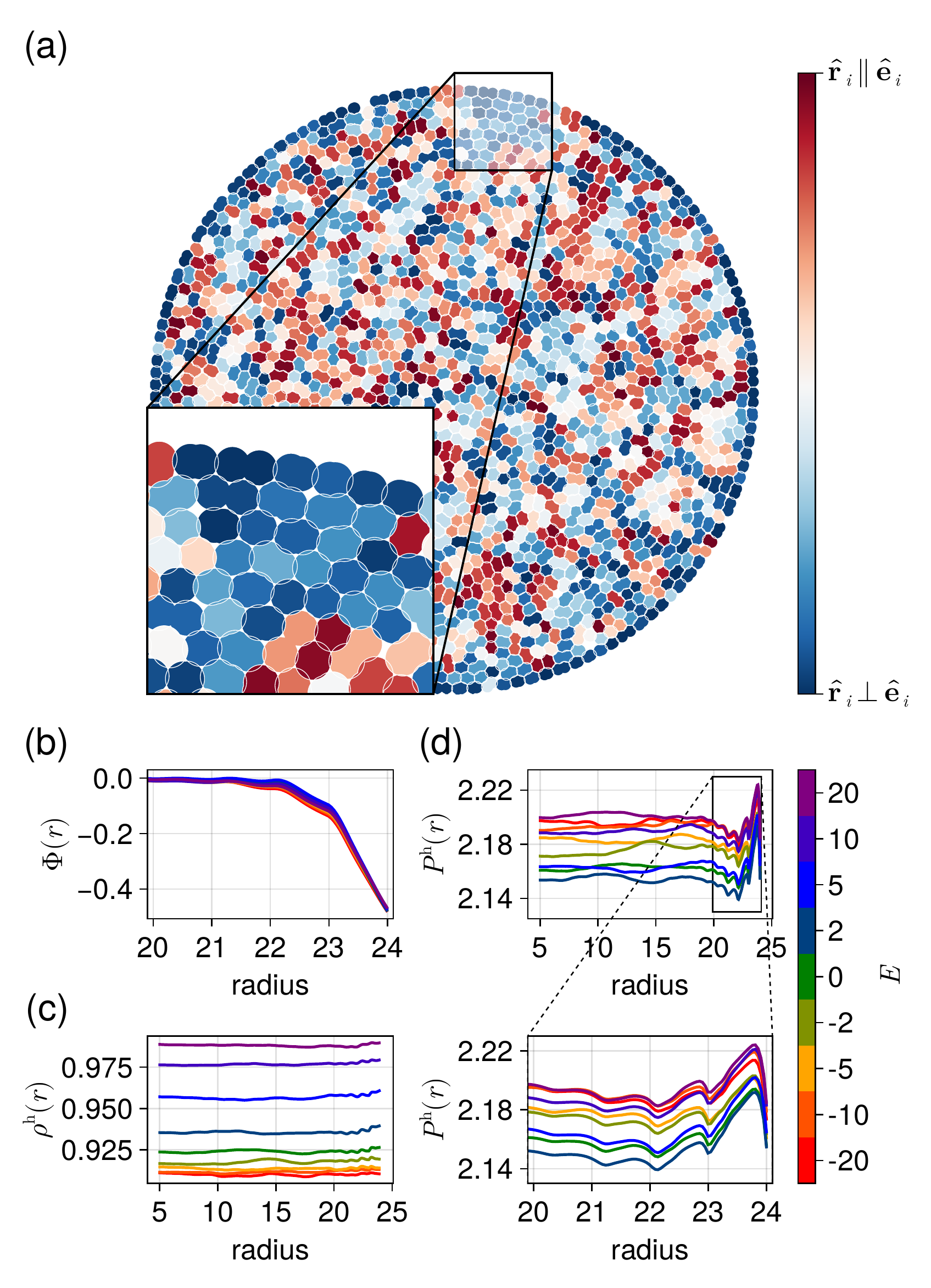}
	\caption{Spatial homeostasis measurements in annuli of equal area.
        \panel{pan:hom_snapshots}~Snapshot of homeostasis simulation with $E=-20$, highlighting polar cell orientation (radial: red, tangential: blue). Cells form highly ordered layers near the confining wall.
        \panel{pan:hom_radial_order}~A nematic order parameter captures local structures of the cell populations, showing tangential alignment at the wall, which propagates into the bulk.
        \panel{pan:hom_radial_activedensity}~Spatial measurement of active density $\rho^\text{h}$ captures the global monotonic increase across the range of degradation factors $E$, but no significant dependence along the spatial coordinate can be observed. 
        \panel{pan:hom_radial_pressure}~The homeostatic pressure shows a non-monotonic behavior approaching the confining wall while also reducing relative differences between populations (see \suppnoteadditionalresults). Below is a zoom in to the previous panel at the region close to the wall.}
	\label{fig:figure5}
\end{figure}

To understand how the confinement affects population properties and the outcome of competition, we perform a spatial analysis of the confined homogeneous system. We start by examining the effect of confinement on cell orientation. Qualitatively, this could already be seen in the snapshots on the right side of \cref{pan:comp_snapshots}, but to demonstrate the effect more clearly, we show in \cref{pan:hom_snapshots} a homeostasis simulation snapshot (for a clone with $E=-20$). Cells are colored by the nematic orientation
\begin{equation}
    \phi_i = 2\left(\hat{\mathbf{r}}_i\cdot\hat{\mathbf{e}}_i\right)^2 - 1
\end{equation}
of their backbone direction $\hat{\mathbf{e}}_i$ relative to radial direction $\hat{\mathbf{r}}_i=\mathbf{r}^\text{cm}_i/||\mathbf{r}^\text{cm}_i||$ at their position, indicating radial (red) or tangential (blue) alignment. Cells touching the wall are highly tangentially ordered, as expected from their anisotropic shapes. The inset shows a zoom-in to a representative region close to the border with dimensions $8$x$8$. It highlights visually up to $7$ ordered layers of cells. To characterize the cell alignment quantitatively, we examine collective nematic order in the polar coordinate system as a function of radial distance from the center, as shown in \cref{pan:hom_radial_order}. Each cell's orientation contributes towards this order parameter $\Phi$ with a weight $w_i$, which is the area fraction of that cell in the considered region (see \suppnoteanalysis). Consequently, the order parameter is the averaged weighted orientation in a set of cells,
\begin{equation}
    \Phi = \frac{\sum_i w_i\phi_i}{\sum_i w_i},
\end{equation}
where $\Phi = +1 \text{ or} -1$ means perfect radial and tangential alignment, respectively, and zero means complete disorder. The visually clear tangential order seen in the snapshot is confirmed near the boundary and also propagates into the bulk for multiple cell diameters, as discussed in \suppnoteadditionalresults. The orientation is shown only for the region near the boundary ($r>20$), since the bulk region is fully disordered with $\Phi\approx 0$, as expected.

Having characterized how the mechanical confinement shapes population structure, we now turn to examine its effect on the active density and pressure, being the principal drivers of competition. In \cref{pan:hom_radial_activedensity}, the spatial profile of the active density $\rho^\text{h}$ is essentially flat, with no considerable dependence on position. For different clones, this homogenous level varies monotonically with the degradation factors $E$, matching the trend in \cref{pan:hom_global_activedensity}. In contrast, the homeostatic pressure $P^\text{h}$ shows a complex spatial profile (\cref{pan:hom_radial_pressure}), which can strongly influence competition outcomes. Notably, the difference in $P^\text{h}$ between clones diminishes towards the boundary, as shown in \suppnoteadditionalresults.

\subsection{Interplay of competition mechanisms}
\label{subsec:PID}

The strong variation of homeostatic pressure shown above suggests that the interplay between altered pressure and active density near the wall jointly shapes the local fitness landscape. To move from a descriptive view towards a more quantitative statement about the relative importance of both competition drivers, we make use of an information-theoretical concept: the partial information decomposition (PID) \cite{chan2017gene}. PID provides intuition by separating how multiple variables contribute to a target quantity. Specifically, it identifies the \textit{unique} information that each source carries about an outcome\footnote{Additionally, it provides information shared redundantly, and the information that emerges only from their joint action.}. In our context, PID therefore offers a way to evaluate how much of the fitness outcome is uniquely attributed to either active density or homeostatic pressure, and therefore the influence of the two mechanisms. We compute PID across all clone-pair combinations simultaneously, treating $(\Delta \rho^\text{h},\ \Delta P^\text{h},\ N_\text{H}/N^\text{t})$ as a time-dependent dataset, which additionally allows us to resolve how the relative contributions of the two mechanisms evolve during the simulations. To establish a baseline for interpreting the confined case, we first analyze PID in the periodic system. Here, we focus on unique contributions. The full analysis of PID, including synergistic and redundant information, is reported in the \suppnoteadditionalresults.

First, we summarize the results for the periodic system. As we see for 1D in \cref{pan:PID_1D_PBC}, the unique contribution from active density differences grows (green), while the information from homeostatic pressure differences is near zero (orange). This is consistent with the interpretation that in 1D, opportunistic competition alone decides about the winning clone. In 2D (\cref{pan:PID_2D_PBC}), the information is similar to 1D on short timescales, while eventually, opportunistic contributions decrease and pressure becomes important again, indicating their joint importance at the simulation end, as discussed earlier. This timescale separation is also found in individual population fraction time series, similar to \cref{pan:comp_global_timeseries}, where non-monotonicities can be observed, favoring the H clone on short and the L clone on long timescales (see \suppnoteadditionalresults).

\begin{figure}[ht]
    \centering
    \includegraphics[width=0.5\textwidth]{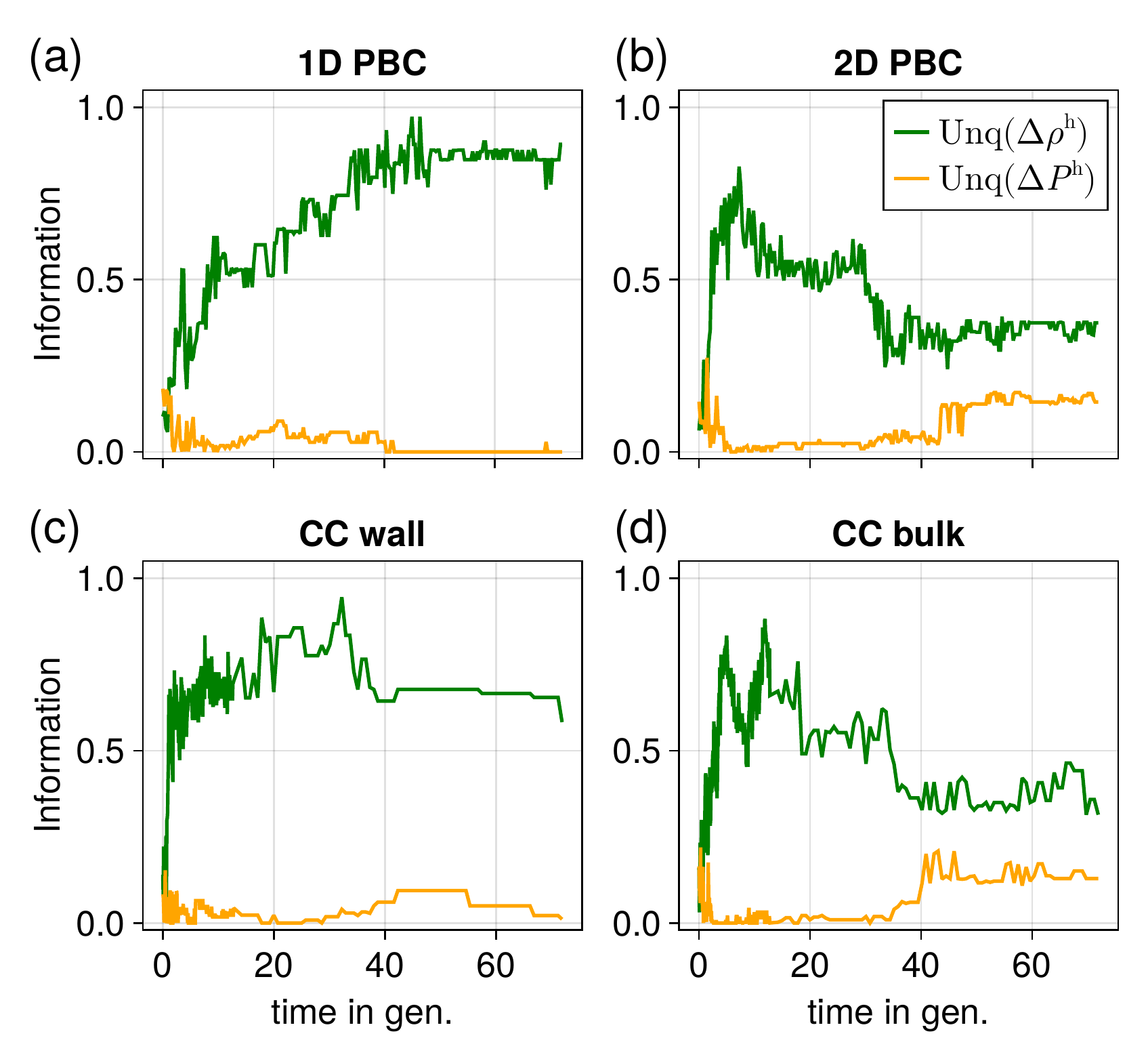} 
	\caption{Unique contributions to fitness of differences in active density $\Delta\rho^\text{h}$ and homeostatic pressure $\Delta P^\text{h}$ calculated by partial information decomposition. Information summarized for cell populations
        \panel{pan:PID_1D_PBC}~in a periodic (PBC) domain in 1D,
        \panel{pan:PID_2D_PBC}~in a PBC domain in 2D,
        \panel{pan:PID_CC_wall}~in a circular confinement (CC) at the confining wall, and
        \panel{pan:PID_CC_bulk}~in CC at the colony center (bulk).}
	\label{fig:figure6}
\end{figure}

Applying the PID analysis to the confined cell populations now puts us in a position to develop a hypothesis on the relation between the spatial order evident in the confined homeostasis populations and the observed change in fitness near the boundary. Since the homeostatic pressure shows a complex behavior along the radial coordinate due to layering effects of the cells, while fitness gradients are mostly monotonic (\cref{pan:comp_radial_fitness_2} and \cref{pan:comp_radial_fitness_20}), we therefore analyze PID separately in the bulk and at the wall.

The PID signals in the colony center closely match the 2D PBC baseline: both active density and homeostatic pressure contributions show similar magnitudes and temporal trends (compare \cref{pan:PID_2D_PBC} and \cref{pan:PID_CC_bulk}). This indicates that competition mechanisms in the bulk of confined colonies operate essentially as in the periodic system. This agreement is expected, since the population in the center is not influenced by the outer wall and exhibits the same disordered structure and pressure levels as in the setting without mechanical boundaries, as shown in Fig.~\ref{fig:figure5} (and \suppnoteadditionalresults for further insights).

However, near the population's edge, the unique information is reminiscent of the 1D case, with homeostatic pressure giving almost no unique contribution to the competition outcome while active density and therefore opportunistic dynamics locally decide about winner and loser (\cref{pan:PID_CC_wall}). We hypothesize that the boundary region behaves as a quasi-1D system: strong alignment and layering make opportunistic, interface-driven competition dominant at early times, similar to true 1D dynamics. However, unlike a strictly 1D interface, the boundary layer exchanges cells with the bulk (see \suppnoteadditionalresults), and this slow radial mixing eventually restores the influence of homeostatic pressure. This explains the cases in which the initial opportunistic dominance gives way to pressure-driven fitness differences at later times, as also observed in scenarios such as \cref{pan:comp_radial_2vs-20}.

\section{Discussion}
\label{sec:discussion}

Cell competition in growing tissues can arise through several mechanisms, including adversarial interactions and homeostatic competition driven by mechanical feedback \cite{basan2009homeostatic}. A key factor identified in earlier work is the rate at which dead cells are cleared: faster clearance enables a larger population of active growing cells and higher likelihood to fill newly freed space leading to a competitive advantage, even when division and death rates are identical. This “opportunistic” mechanism was demonstrated in a 1D model by Pollack et al. \cite{pollack2022competitive}.

Here we extend this picture to two-dimensional aggregates and show that dimensionality and confinement substantially reshape the competitive landscape. In systems without mechanical boundaries, our results recover the 1D prediction: faster-degrading clones reach higher homeostatic density and dominate through the opportunistic mechanism. In two dimensions, however, pressure again plays a stronger role on long time scales, sometimes shifting competition outcomes toward the clone favored by homeostatic pressure differences, in this case the L clone. The origin of this 2D-specific behavior remains open and may relate to the more complex spatial arrangements possible in higher dimensions. 

Introducing a rigid circular boundary adds another layer of geometric structure, since competitive advantages become spatially dependent. Globally losing clones can perform significantly better near the confining wall, sometimes even becoming local winners, while globally dominant clones can lose part of their advantage in these regions.

The partial information decomposition (PID) analysis provides a quantitative view of how the two competition mechanisms shape fitness across geometries. In periodic systems, PID revealed a clear dimensional dependence: in 1D, competition is governed almost exclusively by opportunistic dynamics, whereas in 2D, the relative influence of pressure increases over time. Under mechanical confinement, PID showed that these contributions become spatially dependent: the colony center behaves like a 2D bulk environment, while the boundary region exhibits quasi-1D characteristics with a pronounced early-time dominance of opportunistic competition. Yet, slow radial mixing at the wall eventually restores the role of pressure, demonstrating that the balance between the two mechanisms is not static but time-dependent and shaped by emergent cell flows. Together, the PID results highlight that neither mechanism is intrinsically dominant; rather, their relative importance emerges from the interplay between dimensionality, interface availability, and confinement-driven spatial order. This framework thus offers a mechanistic lens for predicting how tissue architecture and microenvironmental constraints modulate competitive outcomes.

An interesting possibility is that the observed non-monotonicities and time-dependent effects we observe (in population fraction as well as PID) might be related to a coarsening of the system: Although initially well mixed, small clone patches vanish stochastically while bigger clusters grow, even without competition (see \suppnoteadditionalresults). This raises the question of how such coarsening dynamics might interact with the competition mechanisms: since opportunistic competition acts primarily at the interface between populations (Appendix D \cite{pollack2022competitive}), could the shrinking of these interfaces during coarsening reduce its overall influence? On the other hand, homeostatic pressure competition is a bulk effect \cite{basan2009homeostatic}, seemingly independent of interfaces, so might it remain largely unaffected? Similar coarsening occurs in 1D as well, although to a lesser extent (also \suppnoteadditionalresults), and without an obvious shift in the balance between the competition mechanisms. This leads us to ask whether dimensionality plays a more intricate role than anticipated, or whether other mechanisms compensate for the reduced coarsening in lower dimensions.

Our results raise several directions for future investigation. First, extending the present framework to three-dimensional tissues would provide a more complete picture of how dimensionality and spatial constraints shape competition. A 3D geometry would capture architectural features relevant to many in vivo contexts, bridging from epithelia \cite{de2012drosophila} to solid tumors \cite{gerlinger2012intratumor}. Second, a soft or deformable confinement, such as a boundary composed of passive cells from a surrounding tissue \cite{survival_merino_2016,levayer2020solid}, would likely produce weaker spatial order than rigid walls but may still generate spatially heterogeneous fitness patterns. Studying how soft or dynamically remodeled confinement influences the local contributions of pressure and active density would help determine whether the phenomena identified here persist under more physiological conditions. Finally, cell motility offers a natural route to reintroduce mixing \cite{sunkel2025motility} within otherwise structured or coarsened colonies. Active rearrangements can restore or enhance interface-rich configurations, suggesting that motility may shift competition toward the opportunistic mechanism even in settings where static architecture suppresses it. Together, these extensions would broaden the applicability of the information-based approach introduced here and help clarify how real tissues balance mechanical constraints, spatial organization, and competitive interactions.\\

\acknowledgements
We thank Jonas Isensee and Lukas Hupe for their technical support, development, and maintenance on software and the simulation framework \textit{InPartS}, and Christina Goss and David Beine for early contributions towards the model implementation. We thank Stefan Klumpp, Timo Betz, Ulrich Parlitz, Torben Sunkel and Tobias Büchner for valuable discussions, feedback, and suggestions. We thank Ramin Golestanian and the Department of Living Matter Physics at MPI-DS for their support. We acknowledge support from the Max Planck Society as well as the Max Planck School Matter to Life, which is jointly funded by the Federal Ministry of Education and Research (BMBF) of Germany and the Max Planck Society. We acknowledge support also from the Deutsche Forschungsgemeinschaft (DFG, German Research Foundation) – Project-ID 449750155 – RTG 2756. 

\interlinepenalty=10000

\end{document}